\documentclass[preprint]{elsarticle}

\pdfoutput=1
\usepackage{graphicx} % Used to insert images
\usepackage{geometry} % Used to adjust the document margins
\usepackage{amsmath} % Equations
\usepackage{amssymb} % Equations
\usepackage[utf8x]{inputenc} % Allow utf-8 characters in the tex document
\usepackage{hyperref}
\usepackage{booktabs}  % table support for pandoc > 1.12.2
\usepackage{multirow}
\usepackage[T1]{fontenc}
\usepackage[linesnumbered]{algorithm2e}
\makeatletter
\def\ps@pprintTitle{%
 \let\@oddhead\@empty
 \let\@evenhead\@empty
 \def\@oddfoot{Accepted Manuscript. https://doi.org/10.1016/j.jpdc.2016.04.014\hfill}%
 \let\@evenfoot\@oddfoot}
\makeatother

\begin{document}

% Document parameters
\title{The GPU-based Parallel Ant Colony System}
\author[uos]{Rafa{\l} Skinderowicz}
\ead{rafal.skinderowicz@us.edu.pl}

\address[uos]{University of Silesia, Intitute of Computer
Science,\\B\k{e}dzi\'nska 39, 41-205 Sosnowiec, Poland\\
\vspace{1em}
       {\rm
       \textcopyright 2016. This manuscript version is made available under the CC-BY-NC-ND 4.0 license http://creativecommons.org/licenses/by-nc-nd/4.0/
       }
}

\begin{abstract}
The Ant Colony System (ACS) is, next to Ant Colony Optimization (ACO) and the
MAX-MIN Ant System (MMAS), one of the most efficient metaheuristic algorithms
inspired by the behavior of ants.
In this article we present three novel parallel
versions of the ACS for the graphics processing units (GPUs). To the best of our knowledge, this is the
first such work on the ACS which shares many key elements of the ACO and the
MMAS, but differences in the process of building solutions and updating the
pheromone trails make obtaining an efficient parallel version for the GPUs a
difficult task. The proposed parallel versions of the ACS differ mainly in their
implementations of the pheromone memory. The
first two use the standard pheromone matrix, and the third uses a novel
selective pheromone memory. Computational experiments conducted on several
Travelling Salesman Problem (TSP) instances of sizes ranging from 198 to 2392
cities showed that the parallel ACS on Nvidia Kepler GK104 GPU (1536 CUDA cores)
is able to obtain a speedup up to 24.29x vs the sequential ACS running on a
single core of Intel Xeon E5-2670 CPU. The parallel ACS with the selective
pheromone memory achieved speedups up to 16.85x, but in most cases the obtained
solutions were of significantly better quality than for the sequential
ACS.
\end{abstract}


\maketitle

\section{Introduction}
\label{sec:Introduction}

The possibility of using graphics processing units (GPUs) to perform general
purpose computations appeared recently and quickly became a subject of intense
research. GPUs offer attractive performance to energy consumption and the cost
of purchase ratio, and allow to perform many types of computations more quickly
while maintaining the same cost in relation to the CPUs~\cite{a29}.
The usefulness of GPUs is evidenced by the fact they are used in approx.~10\% of the
fastest supercomputers in the world~\cite{top500}.
On the other hand, due to significant differences from the CPU architecture, the
use of GPUs often requires significant changes in the algorithm. GPUs contain a
large number of relatively simple computing units (processing elements), and
therefore taking full advantage of their performance requires an appropriate
division of calculations into subtasks. At the same time, the number of high
latency operations should be minimized, particularly involving global memory
access~\cite{a14}.

Metaheuristic algorithms allow to find approximate solutions of good quality to
many difficult optimization problems in a relatively short period of time. Good
examples of efficient metaheuristic algorithms are algorithms inspired by the
foraging behavior of ants, including the Ant Colony Algorithm (ACO), the Ant
Colony System (ACS) and the MAX-MIN Ant System (MMAS)~\cite{a27}.

Similarly to other metaheuristic algorithms, ant colony algorithms are
computationally demanding, therefore much research effort was put into
developing efficient parallel versions for multi-processor computers~\cite{a20}.
In this article we present novel parallel versions of the Ant Colony System
dedicated to the GPU. To the best of our knowledge, this is the first such work,
although GPU versions of some ant colony algorithms, including the ACO and the
MMAS, were proposed in the literature~\cite{a12, a20}.
The ACS algorithm is very similar to the ACO and the MMAS, however, there are
important differences which make efficient parallelization of the algorithm more
complicated. Specifically, we focus on \emph{the local update} of the pheromone memory,
which has a significant impact on both the algorithm runtime and the quality of
the solutions obtained. We use the Travelling Salesman Problem (TSP) to compare
the performance of the algorithms.

%\subsection{Contributions}
%\label{sec:Contributions}
Based on the parallel versions for the GPU of the ACO and the MMAS as described
in the literature, we propose two parallel versions of the ACS which differ in
the implementation of the pheromone memory update which affects both the
algorithm runtime and the quality of the results. We also present a third
parallel version of the ACS which includes a new version of the \emph{selective
pheromone memory} as inspired by an earlier work~\cite{a32, a31}.

The structure of the article is as follows.
Section~\ref{sec:Related_work} contains a review of the literature on parallel
versions of ant colony algorithms with focus on the implementations dedicated
to the GPU. Section~\ref{sec:acs} describes the sequential ACS and the proposed
parallel GPU versions of the ACS.
Section~\ref{sec:experiments}
shows results and analysis of the computational experiments carried on a few
selected TSP instances, focusing on the comparison between the proposed parallel
versions of the ACS.
Section~\ref{sec:summary} contains a brief summary and possible directions for
further research.

\section{Related work}
\label{sec:Related_work}

In the ACO algorithm (and related algorithms), a population of ants (agents)
work simultaneously on solutions to the problem that is being tackled. In
nature, some species of ants use chemical substances called pheromones as a
method of an indirect communication with other members of the colony. If an ant
finds a food source, it deposits small amounts of the pheromone on the path
leading from the nest to the food source. This pheromone trail attracts other
ants and leads them to the food source. Similarly, in the ACO algorithm a number
of ants iteratively construct solutions to the problem based on additional
knowledge about the problem and on virtual pheromone trails that are deposited
on the solution components. The pheromone trails play the essential role of a
collective memory and are usually referred to as a \emph{pheromone memory}.

Despite the parallel nature of the calculations, the parallelization of ant
algorithms is not an easy task and can be done in many ways. An extensive
overview of the different approaches to parallelization of ant algorithms can be
found in~\cite{a20}. The main criterion for dividing the parallel implementations of
ant algorithms is the number of ant colonies used, i.e. \emph{one} or
\emph{multiple}. Another
criterion is the existence of cooperation between the individuals in a colony
and between the colonies. Cooperation most often involves an exchange of the
best solutions found to date. A single colony can be seen as a~\emph{cell model} in
which the ants belong to overlapping neighborhoods which define network
solutions exchange. In this model there are many pheromone matrices that are
updated by ants assigned to respective neighborhoods. Another more popular
approach is the \emph{master--slave} model, in which a designated process (master)
supervises the work of the slave processes (threads), including gathering
knowledge on the best solution found so far.

Most of the parallel ACO versions dedicated to the CPU apply the multi-colony
approach~\cite{a20}. In this case a single processor (or core) deals with the
calculations for a single colony and the speedup is obtained thanks to
simultaneous calculations for a number of colonies. In most cases, periodic
exchange of the best solution found to date is performed between the colonies.
The exchange is usually performed according to a predefined \emph{communication
topology}.
In~\cite{a1}, a comparison can be found of the various communication topologies
(policies) on the convergence of the MMAS algorithm solving the TSP.
The benefits of the communication strongly depend on whether a local search was
used and on the number of iterations respective to the problem size. For small
problems and a large number of iterations, the benefits of communication were
small or even had a negative effect. In the case of a smaller number of
iterations, an exchange of solutions often improved the quality of the results.
Similar conclusions can be found in the work of Chen et al.~\cite{a2}, in which
proof of the convergence (at infinity) of the parallel ACO is given.

Cecilia et al.~\cite{a10} proposed an efficient parallel (GPU) version of the ACO for
the TSP in which an efficient parallelization scheme (data-parallel) was applied of
the solution construction process. The authors also proposed an efficient
parallel version of the probabilistic roulette wheel selection method used by an
ant when looking for the next node (solution component) to append to a current
partial solution. The new method was called \emph{I-Roulette}. Combined, the
innovations allowed to obtain a speedup of 20x as compared to the sequential
version of the algorithm. The parallel version also improved the quality of the
solutions found in a few cases. Uchida et al.~\cite{a5} proposed a GPU version of the
Ant System (AS) for the TSP. A speedup of 22x was achieved by replacing the
sequential roulette wheel selection method with a parallel method called the
\emph{stochastic trial} based on the \emph{I-Roulette} method. The resulting
speedups increased along with the size of the problem, from 4.51x for the
\emph{d198} TSP instance to 22.11x for the \emph{pr1002} instance. However, for
the largest instance, i.e.  \emph{pcb3038}, the speedup was only 11.87x.

Delevacq et al.~\cite{a12} studied a parallel implementation of the MMAS for the TSP on
the GPU. The authors suggested two approaches - the first with a single ant
colony and the second with multiple ant colonies. Two versions were investigated
under the first approach. In the first version, $ANT_{\rm thread}$, each ant was assigned to a
separate thread, and in the second version, $ANT_{\rm block}$, the calculations for a single ant
were performed by a block of threads. The latter turned out to be much faster,
with speedups reaching 19.47x relative to the sequential MMAS. If using the
local search (3-opt heuristic), the maximum speedup was 8.03x. In~\cite{a13}, based on
the algorithm presented by Cecilia et al.~\cite{a10}, Wei et al. proposed an optimized
ACO transition rule in which the maximum value was used instead of calculating
the exact value of the sum of the probabilities. This resulted in further
acceleration of this phase of the algorithm. However, no information was
provided about the absolute runtime of the algorithm.

The TSP was used in most of the work on the parallel ACO for GPUs because of its
simple definition and a straightforward graph representation. However, other
problems were also considered; for example, in the work of Youness et
al.~\cite{a17}, the parallel MMAS for the Satisfiability Problem (SAT) was presented. The
authors used the data parallel approach in which a block of threads worked on a
single ant's solution. Also, the evaluation of the solutions and the
evaporation of the pheromone were parallelized, thus giving a maximum speedup of
21x relative to the sequential implementation. Lower speedups were noted for the
smallest problems.

The parallel ACO for the GPU to accelerate bi-directional pedestrian movement
simulation was described in~\cite{a18}. A speedup of 18x was achieved while
maintaining the simulation quality comparable to the sequential version. Cano et
al.~\cite{a19} presented Parallel multi-objective Ant Programming for the
classification algorithm using GPUs. Very large speedup values were achieved,
i.e. up to 834x relative to the sequential version. Such large values probably
stem from the nature of the computations, in which most of the time was spent on
the evaluation of the solutions that was easy to parallelize. Not without
significance is the fact that the authors implemented the algorithm in Java
rather than in C, which suggests that the sequential version could be sped up,
thus resulting in smaller relative speedup values~\cite{a14}. The classification task was also
addressed in~\cite{a23}, in which a parallel version of the AntMiner algorithm for the
GPU was about 100x faster than the sequential version.

Fingler et al.~\cite{a22} used the GPU to accelerate calculations of the ACO for the
knapsack problem. The parallel ACO with multiple colonies was proposed, in which
each thread block was responsible for the computations of one ant. The resulting
solutions were of lower quality than generated by the best sequential heuristic,
however, they were found in a much shorter period of time (more than 500 times
faster in some cases).

The multi--colony approach was also used by Bai et al.~\cite{a34} to
speedup the MMAS using the GPUs. The GPU accelerated ACO was successfully used
by Cekmez et al.~\cite{a35} to solve the path planning problem of an Unmanned
Aerial Vehicle (UAV). A survey of recent advances in applying the GPUs to
speedup other metaheuristic algorithms can be found in~\cite{a38, a37, a36}.

%\subsection{Background}
%\label{sec:Background_}

Effective use of GPU computing power requires taking into account the distinct
differences between the CPU and GPU architectures, i.e. the parallel ACO
algorithms for the GPU differ from the parallel implementations dedicated to the
CPU. The CPU version usually uses a number of threads that is equal to the
number of processors, while the effective use of the GPU computing power usually
requires that hundreds or thousands of threads be used. In this paper we present
an approach based on the solutions proposed in~\cite{a9, a6, a12, a3}, which use a single colony
with a single pheromone memory matrix. According to the classification given
in~\cite{a20}, this is a coarse-grained master-slave approach. One of the advantages of
this approach is that the solution search process remains relatively faithful to
that of the sequential ACS.

\subsection{Characteristics of GPU computing}
\label{sec:Charakterystyka_obliczen_GPU}

The use of graphics processors for general purpose computing requires taking
into account the significant differences between the GPU and CPU
architectures~\cite{a14}. CPUs are designed to achieve high-speed processing of a single instruction
stream (or multiple in the case of multi-core chips). To achieve this goal, CPUs
are equipped with complex circuits for branch prediction, thus performing
calculations out-of-order and doing vector computations efficiently (e.g. SSE,
AVX). Moreover, they are also equipped with multi-level cache memories of a size
up to a few megabytes per core. All of these solutions are designed to hide
global memory access latency and to maximize the use of the available computing
power. GPUs, on the other hand, are designed to efficiently perform a large
number of independent parallel computations required in the process of graphics
generation. The relative slowness of the computations for a single pixel is
balanced by the high throughput resulting from a significant degree of data-parallelism.

\begin{figure}
\centering
\includegraphics[width=0.5\linewidth]{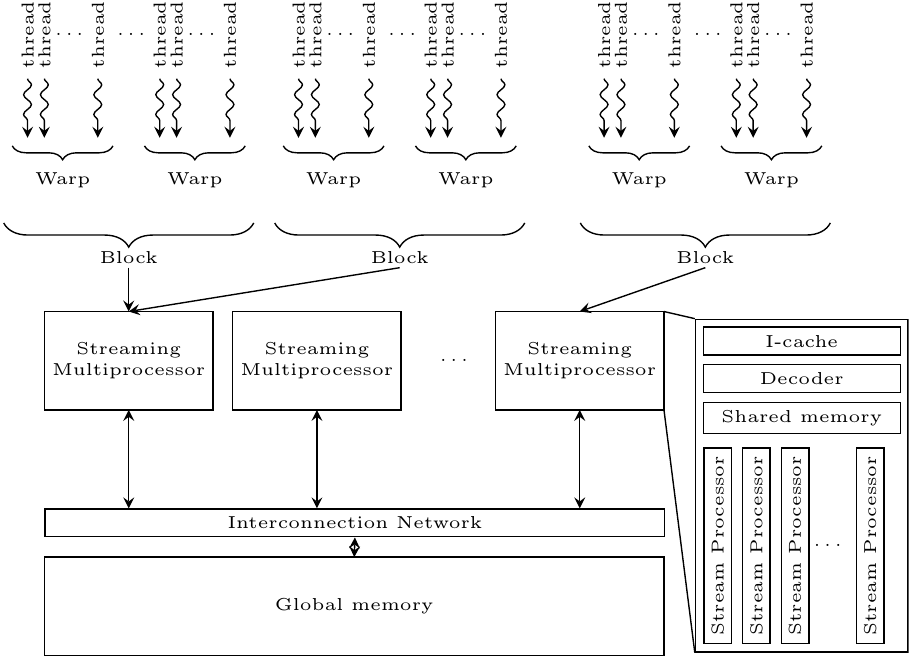}
    \caption{
        GPU architecture overview.
    }
\label{fig:gpu-arch}
\end{figure}

Figure~\ref{fig:gpu-arch} shows a simplified diagram of a GPU. Usually, the GPU
contains a large number of stream processors (or CUDA cores in the case of
Nvidia GPUs), each belonging to one of several streaming multiprocessors (SM).
At any given moment a single core performs calculations for a single thread.
Threads are grouped into \emph{blocks}, with each block assigned to a single SM.
Cores belonging to the same SM share, among others, the register file, local
memory, instruction fetch and decoding, and load/store units. By sharing various
auxiliary units, more computing cores can be packed into a single SM at the
expense of limited flexibility of calculations of individual cores.

The long delays, often at an order of hundreds of cycles, in
accessing the global (main) memory are one of the main obstacles to efficient
parallel computations on GPUs~\cite{a24, a14}. The GPU memory bus is wider than the memory
bus of the CPU and has a relatively large bandwidth (often hundreds of GB/s),
but it is often still not enough to provide data for all of the cores of the
GPU. For this reason, the GPU programming model assumes the use of a large
number of threads, e.g. often tens or hundreds of thousands, between which
switching is fast or even free (zero-overhead scheduling). While a group of
threads is waiting for the completion of data transfer to or from the global
memory, it is possible to perform calculations by threads for which data have
already been transferred. Summarizing, a large number of the GPU processing
elements allows to obtain high speedups, provided that the computations are
largely independent and enough data is transferred in time~\cite{a14}.

\section{Ant Colony System}
\label{sec:acs}

The Travelling Salesman Problem (TSP) is one of the most often studied
combinatorial optimization problems~\cite{a25}. It is NP-hard but has a relatively
simple definition, which allows one to focus one's attention on the algorithm,
hence the TSP has become the de facto standard problem used to test heuristic
algorithms. In this paper we also chose the TSP (symmetric version) as a tool to
test the proposed algorithms, but the conclusions drawn should be useful in the
context of other problems that can be solved by the ACS.

In the TSP the salesman has to visit each city from a given set exactly once and
return to the starting city by taking into account that the trip between any
pair of cities is associated with a given cost. The solution to the problem is a
route with the minimum total cost. TSP can be defined by using a complete graph
$G=(V,A)$, where $V = \{1, 2, \ldots,
n\}$, is a set of nodes (cities), whereas $A = \{ (i, j) | i, j \in V, i \ne j
\}$ is a set of edges connecting the nodes.
A cost $d_{ij}$ is defined for each edge $(i,j) \in A$.

The TSP was one of the first problems used to test the performance of algorithms
as inspired by the behavior of ants, including the Ant System and its enhanced
version - the Ant Colony System~\cite{a15, a26}. In each iteration of the ACS, each ant
builds a complete solution to the problem as follows: each ant starts building
the solution from a randomly selected node (city). In each successive step the
ant extends its current partial solution with one of the previously unvisited
neighboring nodes of the current node (in the case of the TSP each node is
adjacent to all other nodes).  Choosing a node in graph $G=(V,A)$ is equivalent to
choosing a corresponding edge because only a single (unique) edge exists between
each pair of nodes. The choice of the next node $j$ of an ant $k$ located in
node $i$ is carried out according to the formula:
\begin{equation}
j =\begin{cases}
    \arg \max_{l \in J_k^i } [\tau_{il}] \cdot [\eta_{il}]^\beta, &
\mbox{ if }q \le q_0 \\
J,  & \mbox{ if }q > q_0,
\end{cases}
\label{eq:1}
\end{equation}
where $\eta_{il} $ is the cost associated with the selected edge $(i, l)$,
$\tau_{il} $ is the value of pheromone on the edge $(i, l)$, $J_k^i$ is the set
of unvisited (candidate) nodes of ant $k$ and $q_0$ is a parameter.
$J$ is a node (city) selected with the probability defined by:
\begin{equation}
   P(J | i) = \frac{ [\tau_{iJ}] \cdot [\eta_{iJ}]^\beta  }{ 
               \sum_{l \in J_k^i} [\tau_{il}] \cdot
               [\eta_{il}]^\beta }.
    \label{eq:2}
\end{equation}

The choice defined by formula~(\ref{eq:1}) is controlled by parameter $q_0$
($q_0 \in [0,1]$), whose value is
compared with $q$ chosen randomly from uniform distribution. If $q < q_0$, the choice is
greedy, i.e. the edge $(i,j)$ with the highest product of the pheromone trail, $\tau_{ij}$,
and the additional (heuristic) information about the problem, $\eta_{ij}$, is selected.
Otherwise $q > q_0$, and the selection is made randomly with the probability as defined
by Eq.~\ref{eq:2} derived from the Ant System~\cite{a26}. In the first case we can talk about
\emph{exploitation} of knowledge accumulated by the algorithm in the pheromone trail
values. In the second case we can talk about \emph{exploration} of the solution
space~\cite{a15}.

Usually, a relatively high value of $q_0$ is used, e.g. 0.9, so the ACS can find good
quality solutions in less time than the ACO. In order to further reduce the
computation time required to evaluate Eq.~\ref{eq:1}, where $q < q_0$, the choice is narrowed down
to the so-called \emph{candidate set} containing only a subset of the closest unvisited
nodes of the current node. The size of the candidate set, $cl$, is typically 10 to
25~\cite{a15, a27}. The use of the candidate set is justified by the insight that solutions
of good quality are comprised mainly of edges connecting nodes in the direct
vicinity. If all elements of the candidate set are already elements of the
partial solution, the next element is selected from the remaining nodes
(cities). The candidate set for each node (city) is typically calculated at the
beginning of the algorithm and does not change. It is worth mentioning that
Randall and Montgomery~\cite{a28} proposed dynamic update schemes for candidate sets
for the TSP and the Quadratic Assignment Problem. The dynamic version of the
candidate sets allowed to slightly improve the quality of the solutions
relative to the static version, but at the expense of a significantly increased
computation time of the algorithm.

Two types of pheromone updates are used in the ACS. The first, called the \emph{local
pheromone update}, involves "evaporating" a small amount of the virtual pheromone
from the pheromone trail every time the corresponding edge, $(a,b)$, is selected
by an ant. The update is according to the formula:
\begin{equation}
\tau_{ab} \leftarrow (1 - \rho) \cdot \tau_{ab} + \rho \cdot \tau_0 \; ,
\end{equation}
where $\rho \in (0, 1)$ is a parameter and $\tau_0$ is the initial pheromone
trail value. The local pheromone update reduces the likelihood of the same edge
being chosen by subsequent ants, and thus improves exploration of the solution
space.  After all the ants have built complete solutions, a \emph{global pheromone
update} is performed according to the formula:
\begin{equation}
    \tau_{ab} \leftarrow (1 - \alpha) \cdot \tau_{ab} + \alpha \cdot \Delta \tau_{ab} \; ,
    \label{eq:global_update}
\end{equation}
where
\begin{equation*}
    \Delta \tau_{ab} =
    \begin{cases}
        (L_{\rm gb})^{-1}, & {\rm if} \; (a,b) \in {\textit{global\_best\_tour}} \; , \\
    0,                 & {\rm otherwise} \; ,
    \end{cases}
\end{equation*}
where $\alpha \in (0, 1)$ is the coefficient of evaporation (decay) of the
pheromone and $L_{\rm gb}$ is the length of the globally best solution found so far.
The global pheromone update places emphasis on exploitation of the search space
around the best solutions found to date, hence improving the convergence of the
algorithm~\cite{a27}.

\begin{figure}

\begin{algorithm}[H]
\SetKwFor{Parfor}{for}{do in parallel}{end}
\For{ i $\leftarrow$ 1 \KwTo $\#{\it iterations}$  }{
\For{ j $\leftarrow$ 1 \KwTo $\#{\it ants}$ }{
$u$ $\leftarrow$ $\mathcal{U}\{1, \#{\it nodes}\}$  \tcp*[f]{  Choose the first node randomly }

        $route_{{\rm Ant}(j)}[1] \leftarrow u$
}

\For(\tcp*[f]{  Build complete solutions }){ k = 2 \KwTo $\#{\it nodes}$ }{
\For{ j $\leftarrow$ 1 \KwTo $\#{\it ants}$ }{
            $u$ $\leftarrow$ select\_next\_node($route_{{\rm Ant}(j)}$)

            $route_{{\rm Ant}(j)}[k] \leftarrow u$

            local\_pheromone\_update($route_{{\rm Ant}(j)}[k-1], route_{{\rm Ant}(j)}[k]$)
}
}
\For(\tcp*[f]{  Local update on the closing edges }){ j $\leftarrow$ 1 \KwTo $\#{\it ants}$ }{
    local\_pheromone\_update($route_{{\rm Ant}(j)}[\#{\it nodes}], route_{{\rm Ant}(j)}[1]$)
}

    $local\_best$ $\leftarrow$ select\_best($route_{{\rm Ant}(1)}, route_{{\rm Ant}(2)}, \ldots,  route_{{\rm Ant}(\#{\it ants})}$)

\If{ is\_better($local\_best$, $global\_best$)  }{
        $global\_best$ $\leftarrow$ $local\_best$
}

    global\_pheromone\_update($global\_best$)
}
\end{algorithm}
\caption{Ant Colony System.}
\label{alg:1}
\end{figure}

Figure~\ref{alg:1} presents a pseudocode for the ACS. Each ant starts building a solution
from a randomly selected node (line 4). In the subsequent steps the ant extends
the partial solution (lines 6--12) with nodes selected in accordance with formula~(\ref{eq:1}).
The complexity of the algorithm equals
$O( \#{\it iterations} \cdot \#{\it ants} \cdot \#{\it nodes}^2)$
as the number of iterations needed
to select the next node (line 8) is on the order of $O(\#{\it nodes})$. It is often assumed
that $\#{\it ants} = O(\#{\it nodes})$, thus the complexity of the algorithm is 
$O(\#{\it iterations} \cdot \#{\it nodes}^3)$.
The pseudocode for the procedure of the next node selection, in accordance with
formulas~(\ref{eq:1}) and~(\ref{eq:2}) is shown in Fig.~\ref{alg:2}.
In the first place a node is selected from the \emph{candidate set}
containing \emph{cl} nearest neighbors (lines 5--9) of the current node.  Depending on
the value of parameter $q_0$, the selection is made in a deterministic manner (line
13) or randomly with the probability defined by formula~(\ref{eq:2}). If all of the nodes
in the candidate set are already a part of the solution, the next node is chosen
greedily from the remaining nodes (line 18).

\begin{figure}

\begin{algorithm}[H]
\SetKwFor{Parfor}{for}{do in parallel}{end}
\textbf{procedure} select\_next\_node($route_{{\rm Ant}(j)}$)

$cand\_set \leftarrow \{\}$ 

$k \leftarrow $ size($route_{{\rm Ant}(j)}$) \tcp*[f]{  Number of nodes }

$c$ $\leftarrow$ $route_{{\rm Ant}(j)}[k]$  \tcp*[f]{  Current node }

\ForEach{ $e \in nearest\_neighbours[c]$ }{
\If{ $e \notin route_{{\rm Ant}(j)}$  }{
        $cand\_set \leftarrow cand\_set \cup \{e\}$ 
}
}

\eIf(\tcp*[f]{  Select from the closest neighbors }){ $cand\_set \ne \emptyset$  }{
    $q$ $\leftarrow$ $\mathcal{U}(0, 1)$

\eIf(\tcp*[f]{  Greedy selection }){ $r < q_0$  }{
        $next \leftarrow arg\,max_{e \in cand\_set} pheromone[c, e] \cdot heuristic [c, e]$
}{
        $next \leftarrow $ proportional\_selection($cand\_set$)
}
}(\tcp*[f]{  Select from the remaining         }){
    $next \leftarrow arg\,max_{e \in unvisited(route_{{\rm Ant}(j)})} pheromone[c, e] \cdot heuristic [c, e]$
}

$route_{{\rm Ant}(j)}[k+1] \leftarrow next$
\end{algorithm}

\caption{Pseudocode for the procedure of the next node selection in the ACS.}
\label{alg:2}
\end{figure}

\subsection{Parallel ACS for the GPU}
\label{sec:par-acs}

When designing the algorithms presented in this work we were guided by the
conclusions and comments presented in the literature on the parallel ACO and the
MMAS for the GPU. Efficient implementation of the ACS for the GPU, however,
requires that one take into account the significant differences in relation to
the ACO and the MMAS. Based on the results and conclusions presented, among
others, in~\cite{a10, a9, a12}, we applied a model of parallelization in which each
ant corresponds to a single thread block. The stream processors (cores) in
modern GPUs are grouped into so-called \emph{warps} working in the SIMT mode (single
instruction, multiple threads)~\cite{a29}. This means that all threads in the warp
execute the same instruction at the same time. In the case of Nvidia GPUs, the
warp size is typically 32. Each divergence (branch) in the control flow between
threads in the warp, e.g. due to a conditional statement, requires that all the
threads execute the instruction for both paths, which may negatively affect the
computation time. This is one of the reasons why the task parallelism model with
a single ant per thread is not well suited for the GPU~\cite{a10}. For the same reason,
the block size should be a multiple of the size of the warp, and such a rule was
adopted in our implementation. The warp size also affects the size of the
\emph{candidate set}. Following~\cite{a9}, we used the value $cl=32$. As a result, all threads
in the warp perform computations required to select the next node according to
formulas~(\ref{eq:1}) and~(\ref{eq:2}).

\subsubsection{Pseudo-random proportional rule}
\label{sec:Pseudolosowa_regula_przejscia}

In the ACS, the solution construction process is carried out according to the
\emph{pseudo-random proportional} rule as defined by formulas~(\ref{eq:1}) and~(\ref{eq:2}).
In most implementations, only the elements of the candidate set are considered when
applying the rule, and only if the set is empty are the remaining nodes
considered.

To construct a candidate set with $cl = 32$  elements (Fig.~\ref{alg:2} lines
5--8), one needs
to check whether the nodes belonging to the list of \emph{cl} nearest neighbors of the
current node are already a part of the solution. The use of an array of boolean
values indicating the membership of a given node in the partial solution allows
us to perform this step in parallel by all warp threads in a constant time
$O(1)$.
If the candidate set is not empty, a random value $q \in [0, 1)$ is drawn (we used the
XORWOW pseudo-random number generator from the cuRAND library).

If $q \le q_0$ then the current solution is extended with a node to which an
edge with the maximum value of the product of the pheromone trail and the value
resulting from the heuristic knowledge about the problem leads.  The selection
of the maximum value is, of course, an example of a \emph{reduction} operation
that can be performed in $O(\log cl) $ steps. Nvidia GPUs starting with Kepler
architecture offer instructions that enable direct data exchange between the
threads of a warp. These instructions give the threads access to values stored
in the registers of other threads in the warp without the usage of shared
memory~\cite{a29}. This allows, among others, for more effective implementation
of the reduction algorithm~\cite{nvidia}. In our implementation of the reduction algorithm
we applied the variants of the {\it \_\_shfl()} instruction, which is available in the
Nvidia GPUs starting from Fermi architecture (compute capability 3.0 and up).

\begin{figure}
	\centering
	\includegraphics[width=0.8\linewidth]{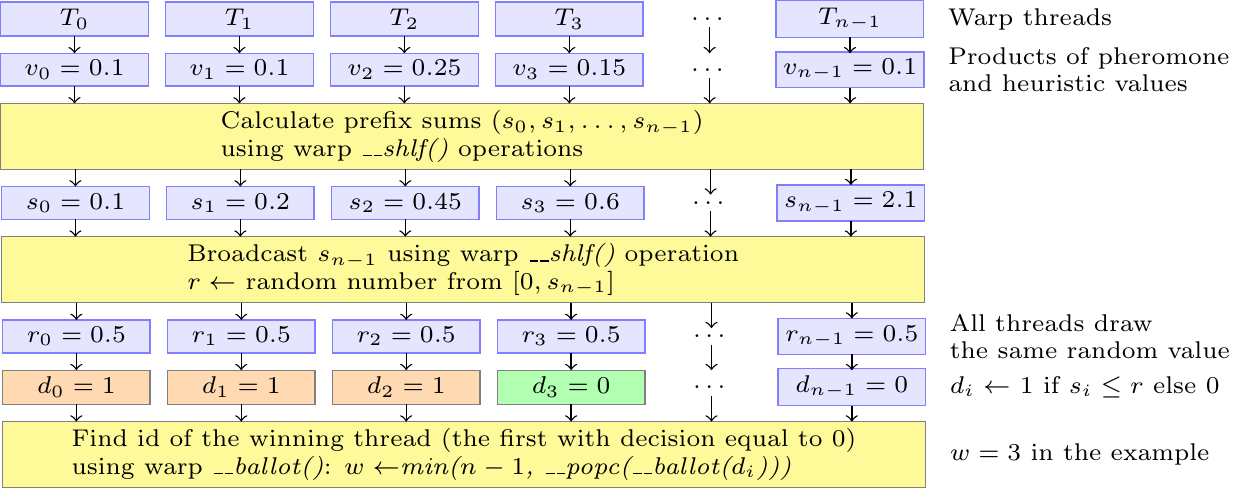}
    \caption{
Example of a parallel warp-level selection of an element in accordance with the roulette
wheel method. In our work we used the Nvidia GPU with 32 threads per warp, i.e. $n = 32$.
    }
	\label{fig:warp-roulette}
\end{figure}

If $q > q_0$ then one of the nodes belonging to the candidate set is chosen with the
probability defined by formula~(\ref{eq:2}), also known as the \emph{roulette
wheel}.
Calculation of the selection probabilities for the elements of the candidate set
requires calculation of the products of the corresponding pheromone trails and
the heuristic knowledge values. Obviously, this can be calculated by using the
reduction algorithm. Figure~\ref{fig:warp-roulette} shows an example of the parallel element selection
according to the roulette wheel method. Because in the ACS the selection is
limited to elements of the candidate set, which in our case had a maximum size
of 32, there was no need for a more complex algorithm, e.g. \emph{I-Roulette} proposed
in~\cite{a10}. Limiting the size of the candidate set to 32 also provides other
advantages. First, there is no need for synchronization barriers because the
threads in a warp work synchronously (SIMT mode). Second, by limiting the size
to 32 elements we could use the efficient intra-warp data exchange functions
introduced in the Fermi architecture which enable threads to access each other's
registers without the need for shared memory. In addition to the {\it \_\_shfl}, we used
the {\it \_\_ballot} function. The result of the function is an $N$ bit integer, with the
$N$-th bit equal to 1 if the \emph{predicate} parameter of the $N$-th thread has a non-zero
value. Because registers are the fastest type of memory available, their use can
have a positive impact on the computation time, as was confirmed by the results
presented in~\cite{a9}.

In the case of an empty candidate set, the next node is one of the remaining
nodes to which the edge with the highest product of the pheromone trail and the
heuristic knowledge values leads. This step can be done in $O(n / p + \log p)$,
where $n$
is the number of the remaining nodes and $p$ is the number of threads in the
block. This is due to the fact that each of the $p$ threads sequentially selects
its own maximum out of the $n/p$ elements, and next the parallel reduction is used
to choose the global maximum from the $p$ thread local maxima. Summarizing,
parallel selection of the next node
in the ACS can be done in $O(n / p + \log p)$, where $n$ is the size of the problem and
$p$ is the number of threads in a block.

\subsubsection{Pheromone update}

Two kinds of pheromone memory updates are performed in the ACS. These are
\emph{global} and \emph{local}. The global pheromone update involves depositing an additional amount
of pheromone on the edges belonging to the best solution found so far. This is
relatively easy to parallelize. In our implementation it is performed in a
separate kernel by a block of 128 threads. Based on the preliminary
computational experiments, we found that a greater number of threads did not
improve the algorithm runtime. This can be explained by the fact that most of
the time in the ACS is spent on constructing the solutions and on local
pheromone memory updates.

The second type of pheromone update in the ACS is called \emph{local} and is
performed after the "transition" of an ant over an edge and after the
corresponding element has been included in the partial solution. It is possible
that the pheromone trail on the edge will be updated more than once in the same
iteration of the algorithm if it is selected by any of the remaining ants. Thus
the local pheromone update is inherent to the solution construction process, and
in our implementation it is carried out in the kernel responsible for the
transition rule.

It should be noted, however, that some differences between the sequential and
parallel search processes could arise.  The obvious difference comes from the
fact that in the parallel version a given number of ants simultaneously
construct solutions to the problem and the relative order of their execution may
vary between successive algorithm runs, while in the sequential version the ants
make their decisions in the same order. If in the parallel implementation a
single kernel execution is responsible for the extension of the ants' partial
solutions with single elements, the basic difference between the sequential and
parallel version boils down to a different relative order of the pheromone
memory updates.  For example, suppose two ants (with indices 0 and 1) are
located at node $a$. In the sequential version the ant with index 0 will select
the next node and will perform the local pheromone update on the chosen edge
before the ant with index 1 starts, thus the action of the former ant could have
an effect on the actions of the latter ant. In the case of the parallel
execution, the ant with index 1 could take action before the ant with index 0
due to slightly different timings of the threads corresponding to both ants.
Moreover, even if the order of execution of the operations on the GPU is the
same as in the sequential version, it is still not certain that ant 1 will see a
new pheromone value written by ant 0. This could result from the \emph{weak} memory
model used in modern GPUs which does not guarantee that changes made in the
global memory by one thread are immediately visible to the other threads and
appear in the order they were made~\cite{a29}.  Actually, the pheromone update
operation is a read-modify-write type of instruction and as such should be
performed \emph{atomically}. The CUDA provides a set of atomic instructions, among them
\emph{atomicCAS} or the compare-and-swap, but their use entails additional overhead and
should be limited.

%But the situation may become more complex.
%Taking the full advantage of the computational capabilities of modern GPUs
%requires that the number of threads (also thread blocks) is much greater than
%the number of physical stream processors (CUDA cores).
%Hence, only a limited number of ants can simultaneously build a solution.

For the purpose of this study, two parallel versions of the ACS for the GPU were
designed (plus one with a selective pheromone memory described in the next
section). In the first
implementation, named \emph{ACS-GPU}, execution of the solution
construction process is close to the execution of the sequential version, i.e.
each ant expands its current solution with a new node and then performs the
local pheromone update. The next step begins after each ant has extended its
partial solutions with single elements. In this version, each step corresponds
to a single GPU kernel execution. The local pheromone update is implemented by
using the \emph{atomicCAS} instruction to prevent any memory inconsistencies.

The second implementation, denoted \emph{ACS-GPU-Alt}, aims to achieve a maximum
speedup, perhaps at the expense of the quality of the solutions found. In this
version the entire solution construction process and the local pheromone update
are performed in a single kernel execution. In this way the number of kernel
calls is reduced from $O(n)$ to $O(1)$, where $n$ is the size
of the problem. Although the computational overhead associated with the kernel
call is quite small (at the order of $\mu s$), it could become significant if the
total algorithm execution time is also small. Another important difference in
relation to the \emph{ACS-GPU} version is resignation from the use of costly atomic
instructions in the local pheromone update process. It may therefore happen, due
to the simultaneous global memory reads and writes, that some of the updated
pheromone values will be lost. Obviously, this could influence the probability
of an edge being selected by subsequent ants, but it should not lead to a
construction of invalid solutions.  It is also worth noting that if the complete
solution is built during a single kernel execution, it is possible that some
ants will build their complete solutions before the other ants even start. This
stems from the fact that there is no enforced synchronization between the
successive steps of the solution construction process, as in the \emph{ACS-GPU}
version. This is especially likely if the number of ants is much greater than
the number of physically available GPU stream processors (CUDA cores), which is
typical in the case of large TSP instances (with thousands of cities).
A detailed list of the proposed algorithms' kernels and blocks/threads configurations
is given in Tab.~\ref{tab:kernels}.

\begin{table}[]
\centering
\small
\caption{List of kernels and blocks/threads configurations of the proposed
GPU algorithms. $\#{\it nodes}$ denotes the size of a TSP instance. The local pheromone
update is performed inside the respective solution construction kernels.}
\label{tab:kernels}
\begin{tabular}{@{}llr@{}}
\toprule
\multicolumn{1}{c}{Algorithm} & \multicolumn{1}{c}{Kernel name} & \multicolumn{1}{c}{Blocks/threads per block} \\ \midrule
\multirow{4}{*}{ACS-GPU}      & \texttt{select\_first\_nodes}             & $1/128$ \\
                              & \texttt{select\_next\_node}              & $\#{\it ants}/32$ \\
                              & \texttt{eval\_ants\_solutions}           & $\#{\it ants}/32$ \\
                              & \texttt{global\_pheromone\_update}       & $1/128$           \\
\midrule
\multirow{2}{*}{ACS-GPU-Alt}  & \texttt{build\_ants\_solutions\_alt}          & $\#{\it ants}/32$ \\
                              & \texttt{global\_pheromone\_update}       & $1/128$ \\
\midrule
\multirow{2}{*}{ACS-GPU-SPM}  & \texttt{build\_ants\_solutions\_spm}     & $\#{\it ants}/32$ \\
                              & \texttt{global\_pheromone\_update}       & $1/128$ \\ 
\bottomrule
\end{tabular}
\end{table}

\subsection{Selective pheromone memory}
\label{sec:Pamiec_selektywna}

Based on our earlier ideas~\cite{a32, a31}, we developed a parallel version of the ACS for
the GPU in which the pheromone matrix was replaced with a \emph{selective pheromone
memory} of a smaller size. In the ACS, the $n \times n$ pheromone matrix holds the
pheromone values for every edge $(u,v)$ of graph $G(V,A)$, whose nodes represent
cities and the edges correspond to roads between the cities (in the case of the
TSP). Every node $u$ has $n-1$ edges connecting it with neighboring nodes, and thus
for every node there are $n-1$ pheromone values stored. In the proposed \emph{selective
pheromone memory} model, for every node only a limited number, $s < n$, of edges can
have a non-minimum value, while the rest are assumed to have a minimum value,
i.e. $\tau_{\rm min}$. In the previous work~\cite{a32, a31}, a slightly different selective pheromone memory
model was proposed in which, similarly, the total number of pheromone values was
limited, but no limit was put on the number of pheromone values for particular
nodes.

The selective memory model is associated with several decisions. It is necessary
to \emph{select} the edges for which the pheromone values should be stored.
Intuitively, the pheromone values should be stored for the "important" edges,
i.e.  for edges which are necessary in order to build solutions of good quality.
Generally, it is difficult to know \emph{a priori} which edges are important for good
quality solutions to be found, hence we applied a \emph{dynamic} selection criterion.
It is applied each time a pheromone trail is updated, therefore, it should be
relatively quick to compute in order not to significantly increase the total
algorithm runtime. For this reason we adopted a simple algorithm working as
follows.  If the pheromone update applies to an edge for which the pheromone
value is already stored in the (selective) pheromone memory, this value is
modified similarly to the ACS with a standard pheromone matrix. Otherwise, a new
pheromone trail value is calculated by using the ACS rules but assuming that the
current value of the pheromone trail equals the minimum pheromone level, i.e. $\tau_{\rm min}$.
At the same time, if the maximum number of pheromone trails, $s$, per node is
exceeded, the new trail overrides the least recently added pheromone trail.
Hence the algorithm resembles the least recently used (LRU) heuristic used in
the CPU cache memory controller implementations.

The pseudocode of the \emph{selective pheromone memory} update algorithm is shown in
Fig~\ref{alg:3}. The selective pheromone memory data are stored as a list of records, one
for every node $u$. A record for a node $u$ consists of a pair of vectors and the
variable \texttt{tail}. The first vector of the pair is an \emph{index vector} and serves to
identify the indices of the nodes which are connected with node $u$. The second
vector stores the values of the pheromone trails on the edges identified by the
first vector. The variable \texttt{tail} stores the index of the most recently added
pheromone trail. Knowing the \texttt{tail} value and the size of the vector $s$ allows us
to easily calculate the index of the least recently added value. As the maximum
number of pheromone trails is limited, the pheromone update process resembles
that of a circular buffer.  Both the pheromone read and the update operations
require, in the worst case scenario, checking the contents of the entire index
vector, hence their complexity both in the average and in the worst case
scenario, is $O(s)$. Assuming that size $s$ of the vectors used in the selective
memory is a (small) constant, the complexity becomes $O(1)$. In the parallel
implementation for the GPU, the search for a value in the index vector could be
efficiently done by using the warp voting functions, i.e. \texttt{\_\_ballot} and
\texttt{\_\_shfl}.

\begin{figure}

\begin{algorithm}[H]
\SetKwFor{Parfor}{for}{do in parallel}{end}
\textbf{update\_pheromone}($u, v, \tau$)

\tcp*[h]{  $nodes_{u}[0..s-1]$ - a vector of nodes indices connected with $u$ }

\tcp*[h]{  $pheromone_{u}[0..s-1]$ - a vector of pheromone trails values }

\tcp*[h]{  $tail_u$ - index of the most recently added element }

\eIf(\tcp*[f]{  The trail for $(u, v)$ exists }){ $\exists i \in \{0,1,2,\ldots, s-1\} \; nodes_{u}[i] = v$  }{
$pheromone_{u}[i] = x$  \tcp*[f]{  Update the pheromone trail value }
}(\tcp*[f]{  There is no trail for the edge $(u, v)$ }){
    $tail_u \leftarrow (tail_u + 1) \mod s$

$nodes_u[tail_u] \leftarrow v$  \tcp*[f]{  Overrides a previous value }

$pheromone_u[tail_u] \leftarrow \tau$  \tcp*[f]{ Overrides a previous value  }

}
\end{algorithm}

\caption{Pseudocode of the selective pheromone memory update algorithm.}
\label{alg:3}
\end{figure}

\begin{figure}
\centering
\includegraphics[]{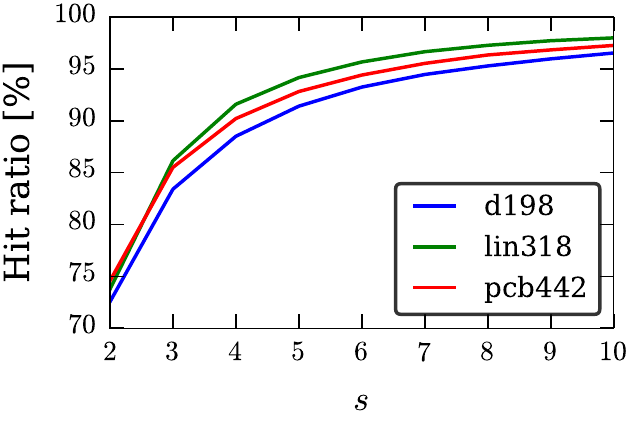}
\caption{
    Each time the local or global pheromone updates are performed in the ACS with the
    \emph{selective pheromone memory} a hit is registered if the corresponding
    pheromone trail's value is already present in the memory structure (line 2
    in Fig.~\ref{alg:3}).
    The plot shows the mean relative ratio of the hits to the total number of
    pheromone updates vs the parameter $s$ which determines the size of the
    memory. The rest of the ACS parameters were consistent with the values
    presented in Section~\ref{sec:experiments}.
}
\label{fig:spm-hit-ratio}
\end{figure}

The selective pheromone memory consists of $n$ records, each containing two
vectors of size $s$. Previous studies have shown that it is enough to remember
only a small number of pheromone trail values to obtain good quality
results~\cite{a32, a31}. 
This is largely consistent with the observations of the convergence of
Population-Based ACO (PACO), in which the pheromone matrix is defined on the
basis of a population of solutions, often of a very small size~\cite{a33}.
Intuitively, it is sufficient that only the edges belonging to the optimal
solution (assuming there is only one) have pheromone trail values that are
different from the minimum value of $\tau_{\rm min}$. Of course, the optimal solution is not
known \emph{a priori}, therefore the selective memory size, i.e. $n \times s$, should be at
least several times larger than the size of the problem, $n$. On the other hand,
the larger the memory size, the higher the computational cost associated with
the pheromone read and update operations. Based on the preliminary experiments,
we chose a value of $s=8$ for which most of the pheromone trails corresponding
to the edges traversed (selected) by ants fit into the memory as shown
in~Fig.~\ref{fig:spm-hit-ratio}.
Hence, at any time no more than 8 edges for every node
could have pheromone trail values different from $\tau_{\rm min}$. Because $s$ is constant, the
final memory complexity of the selective pheromone memory equals $O(ns)=O(n)$,
versus $O(n^2)$ for the standard pheromone matrix.

\section{Experiments}
\label{sec:experiments}

The experiments consist of several parts. We started with a focus on the
parallel ACS implementation by using the standard pheromone matrix for the GPU,
then a performance-oriented investigation of the parameters' values, and we
ended by focusing on the parallel ACS with the selective pheromone memory.

\subsection{Parallel vs sequential ACS}

The first part of the experiments was focused on the ACS algorithm's performance on
the GPU. Three versions of the ACS algorithm were considered. The first, labeled
\emph{ACS-SEQ}, was the reference sequential implementation of the ACS in C language by
Thomas St{\"u}tzle~\cite{acotsp}. The second version was the above-mentioned ACS-GPU, which is
a parallel version of the ACS for the GPU; it is a version in which the process
of solution construction by the ants is closest to the sequential version of the
ACS. The third version, \emph{ACS-GPU-Alt}, is an alternative implementation for the
GPU in which the main emphasis is put on the performance. In this version the
complete solution is built in a single kernel execution and the pheromone memory
updates are performed without the use of atomic instructions, such as CAS.

% \usepackage{booktabs}
\begin{table}[]
\centering
\caption{Specifications of the CPU and GPU used in the computational experiments.}
\label{tab:specs}
\begin{tabular}{@{}lll@{}}
\toprule
                       & CPU                          & GPU                               \\ \midrule
Device                 & Intel Xeon E5-2670 (8 cores) & \begin{tabular}[c]{@{}l@{}}Nvidia Kepler GK104 \\ (1536 CUDA cores $\rightarrow$ 8 SM $\times$ 192 SP)\end{tabular} \\
Clock                  & 2.6 GHz                      & 745 MHz                           \\
L1 cache               & 32 KB                        & 32 KB                             \\
L2 cache               & 256 KB                       & 512 KB                            \\
L3 cache               & 20 MB                        & --                                \\
Memory bandwidth       & 51 GB/s                      & 160 GB/s                          \\
Processing performance & 166.4 GFLOPS                 & 2288 GFLOPS (SP)                  \\ \bottomrule
\end{tabular}
\end{table}

The algorithms were implemented\footnote{The source code is available at
https://github.com/RSkinderowicz/GPUBasedACS} in C++ with the CUDA library (version 6.5) and
compiled using GCC v. 4.8 with the \emph{-O2} optimization switch. The CUDA was used
because of the relative ease of programming the Nvidia GPU that was used in the
experiments. Pseudo-random numbers were generated using the \emph{XORWOW} generator
from the \emph{cuRAND} library. The calculations were performed on a computer equipped
with a CPU and GPU, whose specifications are given in Tab.~\ref{tab:specs}.

The ACS algorithm, similarly to the ACO, requires setting the values of several
parameters. On the basis of the guidelines presented in the literature and on
preliminary computations, the following values of the parameters were used in
the experiments. The number of ants $m$ equals the size of the problem, i.e.
$m=n$ (unless stated otherwise); $\beta = 3$; $\alpha = 0.2$ (global pheromone
evaporation coefficient); $\rho = 0.01 $ (local pheromone evaporation
coefficient).
 For the efficiency of the ACS, the value of the parameter $q_0$ is very important because it
impacts the balance between the exploitation and the exploration of the solution
space. Based on the preliminary experiments, we found that a value of
$q=(N-20)/n$ leads to good quality results in the case of the TSP. It is worth
noting that the high $q_0$ value improves convergence, however, it may lead to getting
stuck in a local optimum. The computations were repeated 30 times for every
combination of an algorithm and set of parameter values.

We agree with Delevacq et al.~\cite{a12} that a reliable evaluation of a parallel
algorithm should take into account the time (speedup) as well as the quality of
the results. The speedup values were calculated in relation to the sequential
implementation of the ACS algorithm from the ACOTSP software by St{\"u}tzle~\cite{acotsp}. A
similar solution was adopted in the literature~\cite{a9, a6} in order to make the
comparisons more reliable and reproducible. The sequential version takes
advantage of only a single CPU core, which makes the GPU vs CPU comparison not
entirely fair, as was pointed out in~\cite{a14}, however, such a convention was adopted
in most of the articles on the parallel GPU versions of the ACO and MMAS
algorithms as presented in the literature, including~\cite{a10, a12}.

% \usepackage{booktabs}
% \usepackage{multirow}
\begin{table}[]
\centering
\caption{
Speedups and mean times of the building ant solutions in the sequential ACS-SEQ
and in the parallel ACS-GPU and GPU-Alt algorithms. The speedup values were
calculated relative to the runtimes of the ACS-SEQ.
}
\label{tab:seq-vs-gpu-speedup}
\begin{tabular}{@{}lrrrrr@{}}
\toprule
\multicolumn{1}{c}{\multirow{2}{*}{Test name}} & \multicolumn{1}{c}{ACS-SEQ}       & \multicolumn{2}{c}{ACS-GPU}                                     & \multicolumn{2}{c}{ACS-GPU-Alt}                                 \\ \cmidrule(l){2-6} 
\multicolumn{1}{c}{}                           & \multicolumn{1}{l}{Time {[}ms{]}} & \multicolumn{1}{l}{Time {[}ms{]}} & \multicolumn{1}{l}{Speedup} & \multicolumn{1}{l}{Time {[}ms{]}} & \multicolumn{1}{l}{Speedup} \\ \midrule
d198                                           & 15.59                             & 7.97                              & 1.96                        & 1.22                              & 12.83                       \\
a280                                           & 35.85                             & 6.72                              & 5.33                        & 2.47                              & 14.52                       \\
lin318                                         & 47.84                             & 7.92                              & 6.04                        & 2.84                              & 16.84                       \\
pcb442                                         & 96.00                             & 14.94                             & 6.43                        & 5.52                              & 17.40                       \\
rat783                                         & 277.71                            & 47.80                             & 5.81                        & 17.64                             & 15.74                       \\
pr1002                                         & 411.91                            & 80.56                             & 5.11                        & 26.39                             & 15.61                       \\
nrw1379                                        & 832.05                            & 159.54                            & 5.22                        & 56.77                             & 14.66                       \\
pr2392                                         & 2,413.45                          & 483.99                            & 4.99                        & 174.51                            & 13.83                       \\ \bottomrule
\end{tabular}
\end{table}

In the first part of the experiments the algorithms ACS-SEQ, ACS-GPU and
ACS-GPU-Alt were run on a set of 8 TSP instances selected from the TSPLIB
repository: \emph{d197, A280, lin318, pcb442, rat783, pr1002, nrw1379, pr2392}.  The
number of iterations of the algorithm was equal to 1000, so the total number of
generated solutions was equal to $1000 \cdot n$, where $n$ is the size of the problem.
Table~\ref{tab:seq-vs-gpu-speedup} shows the mean timings and speedups obtained for the investigated
algorithms relative to the sequential ACS implementation (ACS-SEQ). As can be
seen, the highest speedup was obtained for the ACS-GPU-Alt, which, depending on
the TSP instance, obtained speedups from 12.8 to 17.4 for \emph{d198} and
\emph{pcb442}
instances, respectively. For comparison, the ACS-GPU achieved a minimum speedup
of 1.97 for the \emph{d198} instance and a maximum of 6.4 for the \emph{pcb442} instance.
Small speedup values obtained for the ACS-GPU can be explained by the fact that
the pheromone updates were performed using atomic operations, which are more
expensive than plain read/write operations. The penalty of many more CUDA kernel
invocations in the ACS-GPU is not without significance because each kernel
invocation was responsible for extending the ants' solutions only by a single
element.  In contrast, in the ACS-GPU-Alt, complete solutions were constructed
during a single kernel execution. Moreover, no expensive atomic instructions
were used in the ACS-GPU-Alt pheromone update implementation.

\begin{figure}
	\centering
	\includegraphics[]{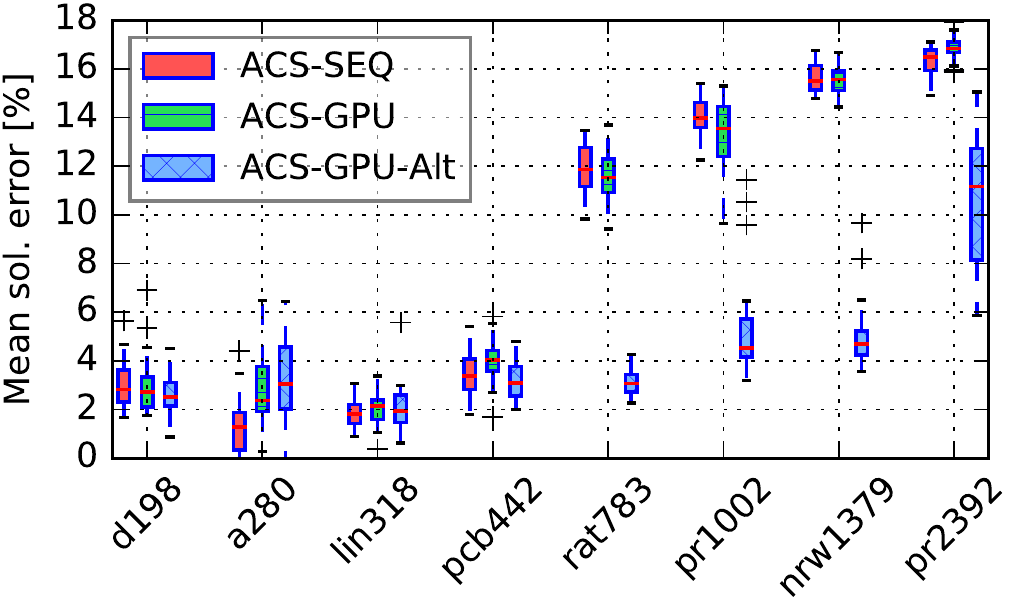}
    \caption{
        Box-plot of the mean solution error (relative to the optimum) for the
        (sequential) ACS-SEQ and the two parallel ACS-ACS-GPU and GPU-Alt
        algorithms
    }
	\label{fig:seq-vs-gpu-quality}
\end{figure}

The question arises as to what extent this "relaxed" model of the ACS execution
affects the quality of the solutions in the case of the ACS-GPU-Alt. Figure~\ref{fig:seq-vs-gpu-quality}
shows the box-plot of the mean distance to the best solution obtained for the
algorithms investigated here. As can be observed, the ACS-SEQ and ACS-GPU
achieved a similar quality of solutions for most of the TSP instances, except
for \emph{a280}, for which the sequential algorithm was clearly better. Much more
interesting is the plot for the ACS-GPU-Alt algorithm, which found solutions of
much better quality for larger instances, i.e. \emph{rat783, pr1002, nrw1379,
pr2392}.
This can be explained by the differences in the implementation of the local
pheromone update. In the ACS-GPU-Alt, the local pheromone updates are not
performed atomically, so some pheromone updates can be lost. Because these
updates result in a reduction (evaporation) of the pheromone on the edges
selected by the ants, the pheromone trail values remain higher than they should.
Limited evaporation of the pheromone causes the solutions built to be less
diverse and more centered around the best solution found so far. We can state
that the process of solution construction in the ACS-GPU-Alt is more
exploitation-oriented, which allows us to obtain better results, especially in
the case of larger TSP instances.

% \usepackage{booktabs}
% \usepackage{multirow}
\begin{table}[]
\centering
\small
\caption{
Mean times $t$ and speedups $s$ required to build $n$ complete solutions for the
ACS-GPU-Alt depending on the period of the local pheromone update (i.e. every 1,
2, 4, 8 and 16 steps).
}
\label{tab:update-freq}
\begin{tabular}{@{}lrrrrrrrrrr@{}}
\toprule
\multirow{3}{*}{Test} & \multicolumn{10}{c}{Pheromone update period}                                                                                                                                                                                                                                            \\
                      & \multicolumn{2}{c}{1}                                  & \multicolumn{2}{c}{2}                                  & \multicolumn{2}{c}{4}                                  & \multicolumn{2}{c}{8}                                  & \multicolumn{2}{c}{16}                                 \\ \cmidrule(lr){2-3} \cmidrule(lr){4-5} \cmidrule(lr){6-7} \cmidrule(lr){8-9} \cmidrule(l){10-11} 
                      & \multicolumn{1}{c}{t {[}ms{]}} & \multicolumn{1}{c}{s} & \multicolumn{1}{c}{t {[}ms{]}} & \multicolumn{1}{c}{s} & \multicolumn{1}{c}{t {[}ms{]}} & \multicolumn{1}{c}{s} & \multicolumn{1}{c}{t {[}ms{]}} & \multicolumn{1}{c}{s} & \multicolumn{1}{c}{t {[}ms{]}} & \multicolumn{1}{c}{s} \\ \midrule
d198                  & 1.1                            & 13.6                  & 1.0                            & 15.0                  & 1.0                            & 15.9                  & 1.0                            & 16.4                  & 0.9                            & 16.6                  \\
a280                  & 2.4                            & 15.0                  & 2.2                            & 16.6                  & 2.0                            & 17.6                  & 2.0                            & 18.2                  & 1.9                            & 18.5                  \\
lin318                & 2.8                            & 17.3                  & 2.5                            & 19.3                  & 2.3                            & 20.4                  & 2.3                            & 21.0                  & 2.2                            & 21.4                  \\
pcb442                & 5.4                            & 17.7                  & 4.9                            & 19.7                  & 4.6                            & 20.9                  & 4.5                            & 21.5                  & 4.4                            & 21.9                  \\
rat783                & 17.4                           & 15.9                  & 15.3                           & 18.1                  & 14.4                           & 19.3                  & 14.0                           & 19.9                  & 13.8                           & 20.1                  \\
pr1002                & 26.1                           & 15.8                  & 22.8                           & 18.0                  & 21.3                           & 19.3                  & 20.7                           & 19.9                  & 20.4                           & 20.1                  \\
nrw1379               & 56.2                           & 14.8                  & 48.2                           & 17.3                  & 43.8                           & 19.0                  & 42.5                           & 19.6                  & 42.0                           & 19.8                  \\
pr2392                & 173.4                          & 13.9                  & 151.2                          & 16.0                  & 136.1                          & 17.7                  & 124.7                          & 19.3                  & 122.9                          & 19.6                  \\ \bottomrule
\end{tabular}
\end{table}

% \usepackage{booktabs}
% \usepackage{multirow}
\begin{table}[]
\small
\centering
\caption{
    Mean distance from the optimum (in \%) for the ACS-GPU-Alt depending on the
    period of performing the \emph{local pheromone updates}. The +/- symbols denote that
    the corresponding result is significantly better/worse than the result for the
    standard ACS in which the local pheromone is performed in every step. The
    significance was calculated using a two-sided nonparametric Wilcoxon rank-sum
    test with the level of significance equal to 0.05
}
\label{tab:freq-quality}
\begin{tabular}{@{}lllllllll@{}}
\toprule
\multirow{2}{*}{\begin{tabular}[c]{@{}l@{}}Pheromone\\ update period\end{tabular}} & \multicolumn{8}{c}{Test}                                                                                                                                                                                                           \\
                                                                                  & \multicolumn{1}{c}{d198} & \multicolumn{1}{c}{a280} & \multicolumn{1}{c}{lin318} & \multicolumn{1}{c}{pcb442} & \multicolumn{1}{c}{rat783} & \multicolumn{1}{c}{pr1002} & \multicolumn{1}{c}{nrw1379} & \multicolumn{1}{c}{pr2392} \\ \midrule
1                                                                                 & 2.655                    & 3.257                    & 2.083                      & 3.226                      & 3.107                      & 5.275                      & 5.013                       & 10.665                     \\
2                                                                                 & 2.794                    & 3.567                    & 1.893                      & 3.042                      & 2.82 +                     & 4.31 +                     & 4.25 +                      & 7.59 +                     \\
4                                                                                 & 3.49 -                   & 3.684                    & 2.057                      & 2.909                      & 2.932                      & 4.27 +                     & 3.83 +                      & 6.95 +                     \\
8                                                                                 & 3.38 -                   & 4.16 -                   & 2.38 -                     & 2.80 +                     & 3.087                      & 4.850                      & 3.94 +                      & 5.19 +                     \\
16                                                                                & 3.23 -                   & 4.05 -                   & 2.348                      & 2.963                      & 3.39 -                     & 4.845                      & 4.37 +                      & 5.81 +                     \\ \bottomrule
\end{tabular}
\end{table}

\subsection{Limiting the number of local pheromone updates}

In order to better investigate the effects of the observed phenomena, the
ACS-GPU-Alt was run with a limited number of local pheromone updates. More
precisely, the local pheromone update was performed for every $k$-th edge selected
by an ant. The following values of $k$ were used: 1, 2, 4, 8 and 16.
Table~\ref{tab:update-freq} contains the mean times of the solution construction depending on the period of
the local pheromone updates $k$. The relative speedup values are also shown.  As
can be observed, the less frequent the updates, the shorter the algorithm
runtime was and thus the higher the speedups. This confirms earlier observations
that the time spent on the execution of local pheromone updates has a
significant impact on the algorithm runtime; for example, for $k=16$ and instance
\emph{pcb442}, a speedup of 21.9 was obtained in contrast to 17.7 if the update was
performed with the standard period ($k=1$). Of course, the greatest speedup could
be obtained by completely removing the local pheromone updates. However, the
local pheromone update is essential to the performance of the ACS in terms of
solution quality.

Table~\ref{tab:freq-quality} shows the mean relative error of the obtained solutions depending on the
period of the local pheromone update. As can be observed, a change in the period
of the local pheromone update has a significant impact on the quality of the
solutions, however, whether the impact is positive or negative depends on the
size of the problem. For the smallest instances, i.e. \emph{d198} and
\emph{a280}, any
increase in the period $k$ of the local pheromone update impairs the quality of
the solutions. For the \emph{pcb442} instance, significantly better solutions were
found for $k=8$, and in the case of the \emph{rat783} TSP instance, the best results were
observed for $k=2$.  For the two largest instances, i.e. \emph{nrw1379} and
\emph{pr2392},
significantly better results were obtained for all $k \ge 2$. However, the best
quality results for the \emph{nrw1379} instance were found for $k=4$, and for
the \emph{pr2392}
instance the best value was $k=8$. These results can be explained by the fact that
the local pheromone update is intended to increase exploration of the problem
solution space. If the local pheromone update is performed less often, the
search process becomes more exploitative, i.e. the solutions constructed are
close to the best solutions found so far. This is beneficial for larger
problems, particularly if the available computational budget is small. On the
other hand, increased exploitation does not improve the quality of the solutions
for the smaller problems, and may even lead to getting stuck in local optima,
which results in worse quality solutions.

Summarizing, a less frequent local pheromone update reduces the computation time
of the parallel version of the ACS for the GPU, however, it affects the quality
of the solutions. For the smaller problems this is disadvantageous, but for
problems of a larger size ($ >1000$ cities) it significantly improves the quality
of the results. Table~\ref{tab:best-sol} shows the best solutions obtained both for the CPU and
the GPU versions of the ACS. As can be observed, in almost every case the GPU
version found better quality solutions and its dominance increased along with
the size of the problem.

\begin{table}[]
\centering
\caption{
Comparison of the best solutions found by the ACS-SEQ and ACS-GPU-Alt. The
figure in brackets is the percentage distance from the optimal solution.
}
\label{tab:best-sol}
\resizebox{\textwidth}{!}{%
\begin{tabular}{@{}lllllllll@{}}
\toprule
\multicolumn{1}{c}{\multirow{2}{*}{Algorithm}} & \multicolumn{8}{c}{Test}                                                                                                                                                                                                           \\
\multicolumn{1}{c}{}                           & \multicolumn{1}{c}{d198} & \multicolumn{1}{c}{a280} & \multicolumn{1}{c}{lin318} & \multicolumn{1}{c}{pcb442} & \multicolumn{1}{c}{rat783} & \multicolumn{1}{c}{pr1002} & \multicolumn{1}{c}{nrw1379} & \multicolumn{1}{c}{pr2392} \\ \midrule
ACS-SEQ                                        & 16046 (1.7\%)            & 2579 (0.0\%)             & 42404 (0.9\%)              & 51695 (1.8\%)              & 9672 (9.8\%)               & 290818 (12.3\%)            & 65018 (14.8\%)              & 434362 (14.9\%)            \\
ACS-GPU-Alt                                    & 15919 (0.9\%)            & 2579 (0.0\%)             & 42203 (0.4\%)              & 51511 (1.4\%)              & 8939 (1.5\%)               & 266416 (2.8\%)             & 58350 (3.0\%)               & 391691 (3.6\%)             \\ \bottomrule
\end{tabular}
}
\end{table}

\subsection{Manipulating the number of ants}

An interesting question is what impact the number of ants $m$ has both on the
runtime and on the quality of the solutions of parallel ACS for the GPU. To
measure this effect, in the following experiments a constant computational
budget, $b$, was set and measured in the total number of solutions built. In a
single iteration of the algorithm, each ant builds one complete solution to the
problem, hence if the number of ants is m and the total number of solutions to
build is b, then the number of algorithm iterations equals $b/m$. It is worth
recalling that in the previous experiments we used the number of ants $m$ equal to
the size of the TSP instance solved, as is often recommended in the
literature~\cite{a10, a9, a12, a27}. The ACS-GPU-Alt algorithm was run for the
two largest instances, \emph{nrw1379}
and \emph{pr2392}, with the number of ants $m \in \{ 128, 256, 512, 1024, n \}$,
where $n$ was the
size of the problem. The smallest number of ants was set to 128 because in order
to make good use of the hundreds (in our case 1536) of GPU CUDA cores, the
number of active threads should be large~\cite{a14, a29}. In our approach a single ant
solution was built by a block of 32 threads, hence the minimum number of threads
was equal to $128 \cdot m$.

Table~\ref{tab:ants-count} shows the results. For both problems the highest mean
speedup and the lowest mean solution error were obtained when the number of ants
was equal to $m=256$. Please note that for both instances, i.e. \emph{nrw1379}
and \emph{pr2392}, the smallest speedups and the largest mean solution error
were obtained for the number of ants equal to the size of the problem, $m=n$.
For example, in the case of the \emph{pr2392} instance, the mean error was
10.67\% for $m=n$, while for $m=256$ the mean error was only 5.29\%.

\begin{table}[]
\small
\centering
\caption{
Comparison of the mean quality of solutions and the mean computation times of
the ACS-GPU-Alt depending on the number of ants, $m$.
}
\label{tab:ants-count}
\begin{tabular}{@{}lrrrrr@{}}
\toprule
Test                     & Ants & Mean error {[}\%{]} & Min. error {[}\%{]} & Time{[}ms{]} & Speedup \\ \midrule
\multirow{5}{*}{nrw1379} & 128  & 4.351               & 3.476               & 51.392  & 16.184  \\
                         & 256  & 4.189               & 3.019               & 51.168  & 16.255  \\
                         & 512  & 4.332               & 3.282               & 51.452  & 16.165  \\
                         & 1024 & 4.289               & 2.949               & 53.955  & 15.415  \\
                         & 1379 & 5.013               & 3.574               & 56.155  & 14.811  \\ \midrule
\multirow{5}{*}{pr2392}  & 128  & 6.174               & 4.209               & 155.085 & 15.559  \\
                         & 256  & 5.290               & 3.374               & 152.309 & 15.843  \\
                         & 512  & 5.631               & 3.904               & 153.194 & 15.751  \\
                         & 1024 & 6.609               & 4.245               & 162.444 & 14.854  \\
                         & 2392 & 10.665              & 5.864               & 173.444 & 13.912  \\ \bottomrule 
\end{tabular}
\end{table}

Summarizing, both the period of the local pheromone update and the number of
ants have a significant impact on the quality of the results. Now the question
is whether combining the less frequent pheromone update with a smaller number of
ants makes it possible to further improve the quality of the solutions and/or
the speedups. In order to answer this question, the ACS-GPU-Alt algorithm was
run for the \emph{nrw1379} and \emph{pr2392} instances with the number of ants
equal to 256 and the period of local pheromone update $k$ equal to 2, 4, 8 and
16. Table~\ref{tab:ants-vs-update-freq} shows the results of the computations. A
larger update period resulted in greater speedups for both TSP instances. The
highest speedups were obtained for $k=16$ and were equal to 20.058x and 18.785x
for the \emph{nrw1379} and \emph{pr2392} instances, respectively. At the same
time, increasing the period of the local pheromone update caused a deterioration
in the quality of the results.  Taking into account the previous results, we
can conclude that both a lower number of ants and less frequent local pheromone
updates can reduce the computation time, but at the expense of the quality of
solutions.

\begin{table}[]
\small
\centering
\caption{
Comparison of the quality of solutions and the computation times of the
ACS-GPU-Alt algorithm depending on the period of the \emph{local pheromone
update}. The calculations were made for the number of ants equal to $m=256$.
}
\label{tab:ants-vs-update-freq}
\begin{tabular}{@{}lrrrrr@{}}
\toprule
\multicolumn{1}{c}{Test} & \multicolumn{1}{c}{Pher. update period} & \multicolumn{1}{c}{Mean error {[}\%{]}} & \multicolumn{1}{c}{Min. error {[}\%{]}} & \multicolumn{1}{c}{Time {[}ms{]}} & \multicolumn{1}{c}{Speedup} \\ \midrule
\multirow{5}{*}{nrw1379} & 1                                       & 4.189                                   & 3.019                                   & 51.168                            & 16.255                      \\
                         & 2                                       & 4.266                                   & 3.321                                   & 45.816                            & 18.154                      \\
                         & 4                                       & 4.223                                   & 3.060                                   & 43.296                            & 19.210                      \\
                         & 8                                       & 4.474                                   & 3.395                                   & 42.103                            & 19.754                      \\
                         & 16                                      & 4.718                                   & 3.464                                   & 41.466                            & 20.058                      \\ \midrule
\multirow{5}{*}{pr2392}  & 1                                       & 5.290                                   & 3.374                                   & 152.309                           & 15.843                      \\
                         & 2                                       & 6.249                                   & 3.413                                   & 137.036                           & 17.608                      \\
                         & 4                                       & 7.077                                   & 3.798                                   & 131.238                           & 18.386                      \\
                         & 8                                       & 8.331                                   & 4.797                                   & 128.694                           & 18.750                      \\
                         & 16                                      & 8.404                                   & 4.801                                   & 126.511                           & 19.073                      \\ \bottomrule
\end{tabular}
\end{table}

\begin{figure}
	\centering
	\includegraphics[]{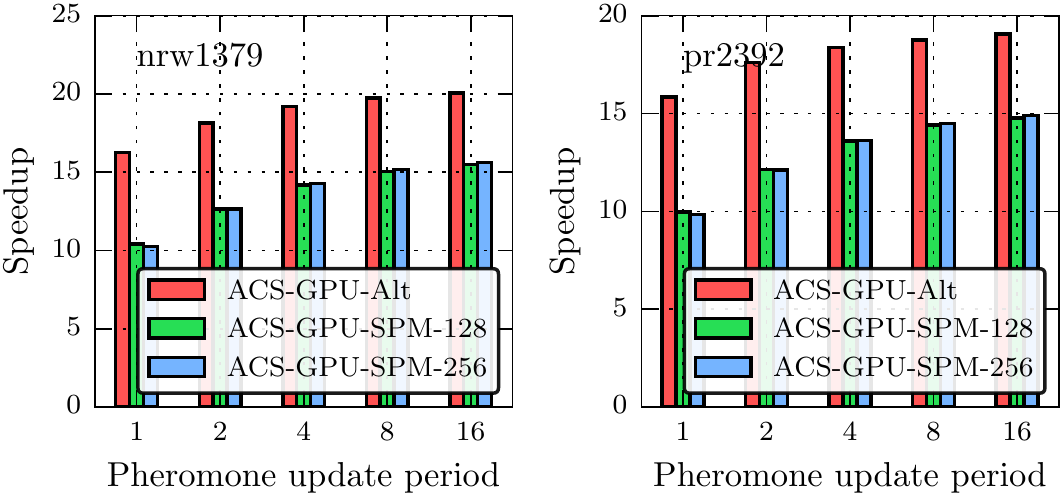}
    \caption{
Comparison of the mean speedups of both the ACS-GPU-Alt and the ACS-GPU-SPM
algorithm depending on the period of the \emph{local pheromone update}. The ACS-GPU-SPM
algorithm was run with the number of ants equal to 128 (ACS-GPU-SPM-128) and 256
(ACS-GPU-SPM-256), while the ACS-GPU-Alt was run with 256 ants. The graph on the
left is for the \emph{nrw1379}, and on the right for the \emph{pr2392} TSP instance.
    }
\label{fig:acs-spm-speedup}
\end{figure}

\subsection{Parallel ACS with a selective pheromone memory}

\begin{figure}
	\centering
	\includegraphics[]{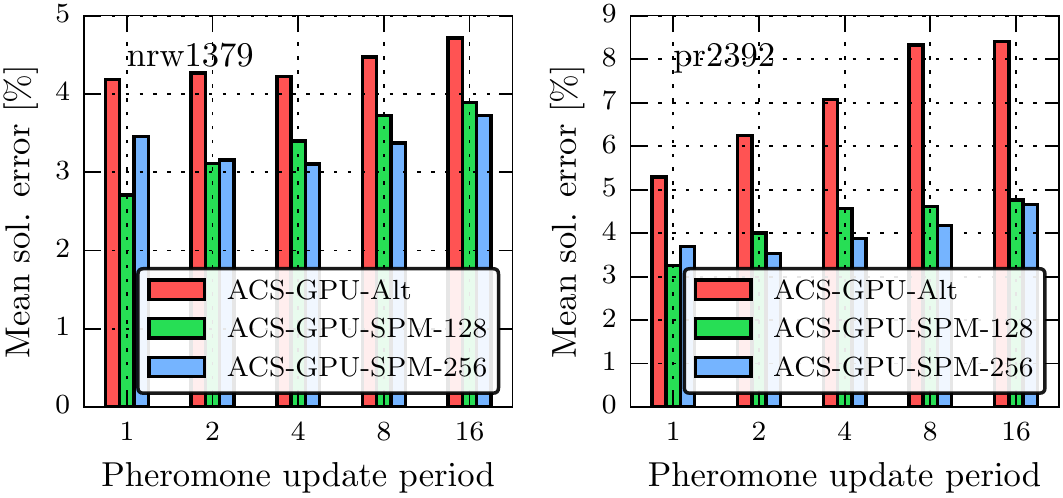}
    \caption{
Comparison of the mean solution error (relative to the optimum) of the
ACS-GPU-Alt and ACS-GPU- SPM algorithms depending on the period of the
\emph{local pheromone update}. The ACS-GPU-SPM algorithm was run with the number
of ants equal to 128 (ACS-GPU-SPM-128) and 256 (ACS-GPU-SPM-256), while the
ACS-GPU- Alt was run with 256 ants. The graph on the left is for the
\emph{nrw1379}, and on the right for the \emph{pr2392} TSP instance.
    }
\label{fig:acs-spm-quality}
\end{figure}

The last part of the experiments was focused on the ACS-GPU-SPM in which the
pheromone matrix was replaced with a selective pheromone memory. The number of
ants and the period of the local pheromone update have the largest effect on
selective memory performance. Based on the results of the previous experiments,
the ACS-GPU-SPM was run with the number of ants $m$ equal to 128 and 256, and
the period of the local pheromone update $ \in \{1,2,4,8,16\}$.
Figure~\ref{fig:acs-spm-speedup} shows the plot of the mean speedup for the
ACS-GPU-Alt and ACS-GPU-SPM algorithms. As can be observed, the speedups for the
algorithm with the selective pheromone memory were lower than for the ACS-GPU-Alt
algorithm with the full pheromone matrix.
This is due to the fact that both the reading and writing operations for the
selective pheromone memory are more costly than for the standard pheromone
matrix implemented using a plain array. The largest differences can be observed
when the local pheromone update was performed in every step of the algorithm
($k=1$). For both instances, the speedups obtained for the ACS-GPU-SPM are about
30\% less than those for the ACS-GPU-Alt. With the increase of the local
pheromone update period $k$, the differences in the algorithms' runtimes
decreased. For $k=16$, the speedup of the ACS-GPU-SPM was approx. 25\% worse than
the speedup of the ACS-GPU-Alt. Although the use of a selective pheromone memory
had a negative impact on the algorithm runtime, it improved the quality of the
results, which can be observed in Fig.~\ref{fig:acs-spm-quality}. In all cases, the mean solutions error
for the ACS-GPU-SPM was much smaller than the error for the ACS-GPU-Alt
algorithm. The largest differences can be seen for the larger of the two
problems, i.e. \emph{pr2392}, and with the increase of the local pheromone update
period. In particular, the ACS-GPU-SPM with the local pheromone update period
equal to 16 obtained a mean error of 3.72\% and 4.66\% for the \emph{nrw1379}
and \emph{pr2392}
TSP instances, respectively. For the same values of parameters the ACS-GPU-Alt
obtained mean error values equal to 4.72\% for \emph{nrw1379} and 8.4\% for the
\emph{pr2392}
instance. This is a typical example of a compromise between the performance and
quality of the results, however, the selective pheromone memory requires much
less space in the global memory of the GPU.

\begin{figure}
	\centering
	\includegraphics[]{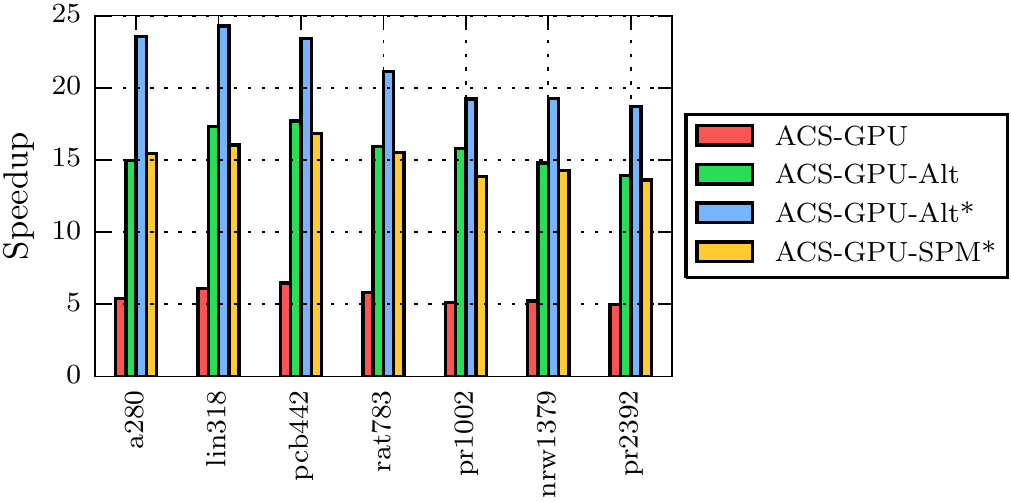}
    \caption{
Comparison of the mean speedups of the parallel algorithms on the GPU for
selected TSP instances. For the algorithms marked with an asterisk (*) the
number of ants was $m=256$ and the period local pheromone update was equal to 4;
in other cases the number of ants was equal to the size of the problem and the
local pheromone update that was performed was equal to 1.
    }
	\label{fig:all-speedup-cmp}
\end{figure}

Figure~\ref{fig:all-speedup-cmp} shows a comparison of the mean speedups for the parallel algorithms
investigated here. The ACS-GPU algorithm was the slowest, for which the average
speedup amounted to a maximum of 6.48x for instance \emph{pcb442}. the The ACS-GPU-Alt
algorithm was much faster, with a mean speedup of approx. 15x and up to 17.72x
obtained for the \emph{pcb442} instance. The largest speedup value of 24.29x was
obtained by the \emph{ACS-GPU-Alt*} algorithm for the \emph{lin318} TSP instance. The
\emph{ACS-GPU-Alt*} version differed from the ACS-GPU-Alt by different values of
parameters: the number of ants $m=256$ and the period of the local pheromone
update $k=4$. These values were chosen based on previous experiments as providing
the best compromise between the speed of execution and the quality of the
results obtained. The ACS-GPU-SPM* version, also with the same values of
parameters as in the ACS-GPU-Alt* version, obtained a speedup that was similar
to the ACS-GPU-Alt, reaching a maximum value of 16.85x for the \emph{pcb442} instance.

In the next part of the experiments, both algorithms, i.e. the ACS-GPU-Alt and
the ACS-GPU-SPM, were run with a fixed time limit, with the number of ants
$m=256$
and the period of the local pheromone update $k=4$. Setting the same time limit
for both algorithms allowed for an easy comparison in terms of the quality of
the solutions. The limit was set for each instance separately based on how much
time it took the ACS-GPU-Alt algorithm to perform 1000 iterations -- however,
with the "base" values of the parameters, i.e. $m=n$ and $k=1$.
Table~\ref{tab:alt-vs-spm} presents a
comparison of the quality of the results. For all instances, except
\emph{lin318}, the
ACS-GPU-SPM produced results of significantly better quality than the
ACS-GPU-Alt, according to the \emph{non-parametric Wilcoxon rank sum test} (the level
of significance was 0.05).  Moreover, in all cases the best solution found by
the ACS-GPU-SPM was of better quality than the solution found by the second
algorithm. In conclusion, despite the fact that the selective pheromone memory
is slower than the pheromone matrix, the improvement in the quality of the
results is so large that it can get ahead of the faster algorithm.

% \usepackage{booktabs}
% \usepackage{multirow}
\begin{table}[]
\centering
\small
\caption{
Comparison of the quality of the results for the ACS-GPU-Alt and ACS-GPU-SPM
algorithms. The \emph{Time limit} column shows the time the algorithms had for a single
execution. Both algorithms were executed with the number of ants equal to
$m=256$
and the period of the local pheromone update equal to 4. The bold font was used
to mark whether the value turned out to be significantly better than for the
other algorithm according to a two-sided, nonparametric Wilcoxon rank sum test
(level of significance 0.05).
}
\label{tab:alt-vs-spm}
\begin{tabular}{@{}lrrrrr@{}}
\toprule
\multirow{2}{*}{Test name} & \multicolumn{1}{l}{\multirow{2}{*}{Time limit {[}s{]}}} & \multicolumn{2}{c}{ACS-GPU-Alt}                                                                     & \multicolumn{2}{c}{ACS-GPU-SPM}                                                                     \\ \cmidrule(lr){3-4} \cmidrule(l){5-6} 
                           & \multicolumn{1}{l}{}                                    & \multicolumn{1}{l}{Mean error {[}\%{]}} & \multicolumn{1}{l}{Best solution}                    & \multicolumn{1}{l}{Mean error {[}\%{]}} & \multicolumn{1}{l}{Best solution}                    \\ \midrule
a280                       & 2.5                                                     & 3.77                                    & \begin{tabular}[c]{@{}r@{}}2619\\ (1.55\%)\end{tabular}   & {\bf 1.50}                              & \begin{tabular}[c]{@{}r@{}}2584\\ (0.19\%)\end{tabular}   \\
lin318                     & 2.84                                                    & 1.89                                    & \begin{tabular}[c]{@{}r@{}}42285\\ (0.61\%)\end{tabular}  & 1.90                                    & \begin{tabular}[c]{@{}r@{}}42162\\ (0.32\%)\end{tabular}  \\
pcb442                     & 5.52                                                    & 2.69                                    & \begin{tabular}[c]{@{}r@{}}51586\\ (1.59\%)\end{tabular}  & {\bf 2.29}                              & \begin{tabular}[c]{@{}r@{}}51242\\ (0.91\%)\end{tabular}  \\
rat783                     & 17.64                                                   & 3.20                                    & \begin{tabular}[c]{@{}r@{}}8968\\ (1.84\%)\end{tabular}   & {\bf 2.51}                              & \begin{tabular}[c]{@{}r@{}}8964\\ (1.79\%)\end{tabular}   \\
pr1002                     & 26.39                                                   & 5.31                                    & \begin{tabular}[c]{@{}r@{}}266657\\ (2.94\%)\end{tabular} & {\bf 3.06}                              & \begin{tabular}[c]{@{}r@{}}264001\\ (1.91\%)\end{tabular} \\
nrw1379                    & 56.77                                                   & 4.09                                    & \begin{tabular}[c]{@{}r@{}}58309\\ (2.95\%)\end{tabular}  & {\bf 3.17}                              & \begin{tabular}[c]{@{}r@{}}57934\\ (2.29\%)\end{tabular}  \\
pr2392                     & 174.51                                                  & 6.05                                    & \begin{tabular}[c]{@{}r@{}}393541\\ (4.1\%)\end{tabular}  & {\bf 4.05}                              & \begin{tabular}[c]{@{}r@{}}388938\\ (2.88\%)\end{tabular} \\ \bottomrule
\end{tabular}
\end{table}

In order to further investigate the performance of the ACS-GPU-SPM algorithm a
few larger TSP instances were selected with the size ranging from 3038 to 14051 cities. 
Table~\ref{tab:bigger} shows the best results out of 5 runs for each instance.
As can be seen, the quality of the results degrades as the size of the problem
grows. It follows from the fact that the size of the solution search space
grows exponentially and the linear increase in the computation time is not
sufficient to keep the quality of the results on a similar level.
The quality of the results could be significantly improved if a local
search heuristic was applied~\cite{a12}.
Nevertheless, the GPU processing power allows the ACS-GPU-SPM algorithm to
generate and evaluate over 300000 solutions per second for the TSP
instance with 14051 cities.

% \usepackage{booktabs}
\begin{table}[]
\centering
\small
\caption{
    Results obtained for the ACS-GPU-SPM algorithm for the selected
    TSP instances of larger size. The algorithm was executed with $m=256$ ants
    and the number of iterations equal to $1000 \cdot n$, where $n$ is the number
    of nodes. The table presents the best results out of 5 runs.
}
\label{tab:bigger}
\begin{tabular}{@{}lrrrr@{}}
\toprule
Test name & \multicolumn{1}{l}{Optimum} & \multicolumn{1}{l}{Best solution} & \multicolumn{1}{l}{Relative error {[}\%{]}}  & \multicolumn{1}{l}{Total time {[}s{]}} \\ \midrule
pcb3038   & 137694                      & 144529                            & 4.96                                         & 252.20                                 \\
rl5915    & 565530                      & 613514                            & 8.48                                         & 1344.87                                \\
pla7397   & 23260728                    & 25179924                          & 8.25                                         & 1957.77                                \\
rl11849   & 923288                      & 1062887                           & 15.12                                        & 8262.88                                \\
usa13509  & 19982859                    & 23641280                          & 18.31                                        & 11863.45                               \\
brd14051  & 469385                      & 535734                            & 14.14                                        & 11730.20                               \\ \bottomrule
\end{tabular}
\end{table}

\section{Summary}
\label{sec:summary}

In the paper, three novel, data-parallel versions of the ACS algorithm for the GPU are
proposed, namely the ACS-GPU, ACS-GPU-Alt and ACS-GPU-SPM. All three algorithms
were implemented by using the Nvidia CUDA framework and experimentally compared
with the reference sequential implementation (ACOTSP) by Thomas
 St{\"u}tzle~\cite{acotsp}. A series of numerical experiments was conducted using a set of TSP
instances of sizes ranging from 198 to 14051 cities chosen from the TSPLIB
repository.

Generally, the solution construction process of the ACS is relatively easy to
parallelize. However, the implementation of the local pheromone trail update is
of key importance to the efficiency of the parallel ACS. The results showed that
the ACS-GPU is closest to the sequential version in terms of solution quality.
It is also the slowest of the three algorithms, with speedups up to 6.43x, mainly
due to the use of the expensive atomic instructions (on the global GPU memory)
during the local pheromone update. The ACS-GPU-Alt algorithm is the fastest, in
which no expensive atomic instructions are used and complete solutions are built
during a single kernel execution, hence diminishing the overhead associated with
the GPU kernel call. The ACS-GPU-Alt achieved a maximum speedup of 17.4x. A
further speedup can be obtained by performing the local pheromone update less
frequently and by using a smaller number of ants. With the number of ants set to
$m=256$ and the local pheromone update performed every 4 steps, the ACS-GPU-Alt
algorithm achieved a maximum speedup of 24.29x for the \emph{lin318} instance.

In the third algorithm, denoted as ACS-GPU-SPM, the
pheromone matrix was replaced with a novel version of the \emph{selective pheromone
memory}. In the pheromone matrix, for every node $u$ there exists a row
of pheromone trails values, one for each edge starting at node $u$.
In the selective pheromone memory, for every node $u$
only a small, constant number, $s$, of pheromone trails values is stored, i.e.
for a \emph{selected} subset of the edges starting at node $u$, while for the rest
the minimum value, $\tau_{\rm min}$, is presumed.
A set of the selected pheromone trails may change
as a result of performing local or global pheromone updates. If the pheromone
trail for the edge is in the selective memory, its value is updated, otherwise
the new pheromone trail completely overwrites the least recently added
pheromone trail to keep the size constant. On the one hand, the selective
pheromone memory causes the solutions that are built to consist primarily of
the selected edges and thus improves the exploitation of the solutions search space.
On the other hand, the rotation of the trails resulting from the pheromone
updates improves the exploration. The parallel implementation of the
selective pheromone memory for the GPU is more complex than the standard
pheromone matrix and turns out to be approx. 30\% slower.

It should be noted, however, that if the ACS-GPU-Alt and the ACS-GPU-SPM were
run with the same time limit, the latter would allow to obtain results of better
quality. Moreover, both the ACS-GPU-Alt and the ACS-GPU-SPM allow to obtain
results of quality better than the sequential ACS; for example, for the
\emph{pr2392} instance, the sequential ACS was able to find a solution with a mean
error of 16.34\% (relative to the optimum) in 2412 seconds, while the parallel
ACS-GPU-SPM found a solution with a mean error of 4.05\% in 174 seconds.

In conclusion, it is possible to achieve an efficient parallel version of the
ACS for the GPU. One of the key elements influencing the algorithm runtime is
the implementation of the pheromone memory, and in particular of the local
pheromone update. Abandoning full compliance with the modus operandi of the
sequential ACS can significantly speed up the algorithm and even improve the
quality of the results.

\subsection{Further research}

As a part of further research, the algorithm should be tested by using a newer
generation of GPUs, and it should be run simultaneously on multiple GPUs.
Perhaps concurrent kernels execution may be used to speed up even further the
computation for multiple ants and multiple colonies~\cite{a39}.
It would also be interesting to see
how convergence of the algorithm changes if a local search heuristic is used,
e.g. 3-Opt, as presented in~\cite{a12}.
\\
\textbf{Acknowledgments:} This research was supported in part by PL-Grid Infrastructure.

\bibliographystyle{plain}
\footnotesize
\bibliography{article}

\end{document}